\documentclass[preprint,showpacs,preprintnumbers,amsmath,amssymb]{revtex4}

\usepackage{graphicx}
\usepackage{dcolumn}
\usepackage{bm}

\def\v#1{\mbox{\boldmath$#1$}}

\begin{document}

\title{Note on the ring approximation in nuclear matter}

\author{E. Bauer}

\email{bauer@fisica.unlp.edu.ar}
\affiliation{
Departamento de F\'{\i}sica, Universidad Nacional de
La Plata and \\
Instituto de F\'{\i}sica La Plata,
CONICET, \\
C. C. 67, 1900 La Plata, Argentina}

\date{\today}

\begin{abstract}
The response function to an external prove is evaluated using
the ring approximation in nuclear matter.
Contrary to what it is usually assumed,
it is shown that the summation of the ring series and the solution
of the Dyson's equation are two different approaches. The numerical
results exhibit a perceptible difference between both approximations.
\end{abstract}

\pacs{21.65.+f, 21.60.-n, 21.60.Jz}

\maketitle

The ring approximation is widely used in many nuclear physics problems.
It consists of the infinite sum of one particle-one hole bubbles, where
the Pauli exchange contribution is neglected~\cite{fe71}. In fact, the ring
approximation is the direct part of the Random Phase Approximation (RPA).
By neglecting the exchange terms, the particle-hole series
reduces itself to a geometric series, which is easily summed up.
As an alternative derivation, it is usually proposed the
ring series as the solution of the
Dyson's equation~\cite{dy49} (see also~\cite{fe71}).
It has come to us as a surprise the existence of an inconsistency between
the interpretation of the ring approximation as a
solution of the Dyson's equation and the explicit evaluation
(employing the Goldstone's rules) of the particle-hole diagrams
which originates the ring series. This inconsistency comes into
play only when one particle-hole configuration is on the mass shell,
for example, when we study the response function.
In this contribution we discuss this point and we show that there
are two alternative approximations. One is the solution of the plain
Dyson's equation and the other is the summation of the ring diagrams.

Let us consider an arbitrary one-body operator ${\cal O}_{\alpha}$ exciting
the nucleus from its ground state $|0>$.
The action of ${\cal O}_{\alpha}$ is characterized by the
nuclear response function per nuclear volume,
\begin{equation}
\label{stfun}
S_{\alpha}(q_0, \v{q}) = -\frac{1}{\pi \, \Omega}\ Im
<0|{{\cal O} _{\alpha}}^{\dag} G(q_0) {\cal O} _{\alpha} |0>,
\end{equation}
where $q_{0}$ and $\v{q}$ are the energy and momentum transfer,
respectively and $\Omega$ is the nuclear volume.
The nucleon propagator
$G(q_0)$ is given by,
\begin{equation}
\label{pprop}
G(q_0) = \frac{1}{q_0 - H + i \eta}\ -
\frac{1}{q_0 + H + i \eta}\,
\end{equation}
being $H$ the nuclear Hamiltonian.  As usual $H=T+V$, where
$T$ is the kinetic energy and $V$ is the residual interaction.
The identity is expressed as,
\begin{equation}
\label{ident}
I = \sum_{n} |n><n|,
\end{equation}
where $|n>$ represent a compleat set of orthonormal states of $H$.
By inserting the identity twice in Eq.~(\ref{stfun}), we have,
\begin{equation}
\label{stfun2}
S_{\alpha}(q_0, \v{q}) = \frac{1}{\Omega} \, \sum_{n} \;
|<n| {\cal O} _{\alpha} |0>|^{2} \, \delta(q_0-q_n),
\end{equation}
where $q_n \equiv E_n - E_0$ ($\hbar = c = 1$),
and $E_n$ are the excitation energies of the eigenstates
$|n>$.

\begin{figure}[h]
\centerline{\includegraphics[scale=0.53]{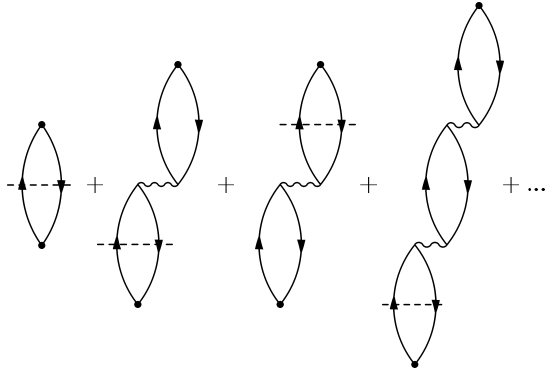}}
\caption{Goldstone's diagrams representing the firsts contributions to
the ring approximation. In each diagram an up (down) arrow constitutes
a particle (hole), a wavy line is the residual interaction and a dot
stand for the external operator. It has been added an horizontal
dashed line to indicate the configuration on the mass shell.}
\label{fig1}
\end{figure}

In the present contribution we explore three different external proofs,
\begin{equation}
\label{extop}
{\cal O}_{\alpha} = \sum_{j} \; e^{i \v{q} \cdot \v{x}_j}
\widetilde{{\cal O}} _{\alpha \, (j)}
\end{equation}
with $\v{x_j}$ denoting the intrinsic coordinate for individual nucleons
and the sum $j$ runs over all nucleons. In
this equation, $\widetilde{{\cal O}} _{C \, (j)}=1$,
$\widetilde{{\cal O}} _{L \, (j)}= \v{\sigma}_{(j)} \cdot \v{\hat{q}} \, \tau_{z \, (j)}$ and
$\widetilde{{\cal O}} _{T \, (j)}= \v{\sigma}_{(j)} \times \v{\hat{q}} \, \tau_{z \, (j)}$ which
are usually named as the isoscalar central and isovector spin-longitudinal and
spin-transversal operators, respectively. The next step is to propose a model for the
Hamiltonian. The simplest choice is to keep only the
kinetic energy $T$. In this case, the final state is a one particle-one hole excitation as the
one drawn in the first diagram in Fig.~\ref{fig1}.
Before we show the response function, in
Appendix~\ref{append},
it is presented the lowest-order polarization insertion $\Pi^{0}(q_0, \v{q})$, which
is further expressed as the sum of it real and imaginary parts,
\begin{equation}
\label{polpro2}
\Pi^{0}(q_0, \v{q}) \equiv {\cal R}(q_0, \v{q}) + \imath {\cal I}(q_0, \v{q})
\end{equation}
Now the response function to ${\cal O}_{C}$, is,
\begin{equation}
\label{stfunc}
S^{0}_{C}(q_0, \v{q}) = {\cal I}
\end{equation}
Next, the residual interaction $V$
is incorporated. As a model for $V$, we use
the one given in Eq.~(\ref{inter}), which can be rewritten as a sum of
a isoscalar central and isovector spin--longitudinal and
spin--transversal terms (see Eq.~(\ref{inter2})).
One way of taking care of the residual interaction is by means of
the Dyson's equation, where a higher-order polarization insertion is obtained
as,
\begin{equation}
\label{dyson}
\Pi^{Dys}(q_0, \v{q}) = \Pi^{0}(q_0, \v{q}) + \Pi^{0}(q_0, \v{q}) \,
 V(q) \, \Pi^{Dys}(q_0, \v{q}).
\end{equation}
In this equation the Pauli exchange terms have been already neglected. For this
reason, this is an algebraic equation which solution is
the sum of a geometric series in $V \, \Pi^{0}$,
\begin{equation}
\label{dyson2}
\Pi^{Dys} = \frac{\Pi^{0}}{1 - V \, \Pi^{0}}.
\end{equation}
Using the solution of the Dyson's equation a new response function
is obtained. This is done by replacing ${\cal I}$
in Eq.~(\ref{stfunc}) by $Im \Pi^{Dys}$,
\begin{equation}
\label{stfuncdy}
S^{Dys}_{C \, (L,T)}(q_0, \v{q}) = \frac{\cal I}
{(1-{{\cal V}_{C \, (L,T)} \, \cal R})^{2}+ ({\cal V}_{C \, (L,T)} \, {\cal I})^{2}}.
\end{equation}

The Eq.~(\ref{dyson2}) is usually interpreted as the sum of a
series of one particle-one hole bubbles. The firsts terms to this
series are shown in Fig.~\ref{fig1}. In these diagrams, a horizontal
dashed-line indicates that this configuration is on the mass shell.
It is interesting to analyze each term in the
series separately, we expand Eq.~(\ref{dyson2}),
\begin{equation}
\label{dyson3}
\frac{\Pi^{0}}{1 - V \, \Pi^{0}} = \Pi^{0}(1+V \, \Pi^{0}+(V \, \Pi^{0})^{2} + \ldots ),
\end{equation}
by taking the imaginary part of this sum, we have,
\begin{eqnarray}
\label{ring6}
Im(\Pi^{0}) & = & {\cal I} \nonumber \\
Im(\Pi^{0} V  \Pi^{0}) & = & 2  {\cal R} \; V  \; {\cal I}\nonumber \\
Im(\Pi^{0}(V \, \Pi^{0})^{2}) & = & (3  {\cal R}^{2} {\cal I} - {\cal I}^{3}) \; V^{2} \nonumber \\
& ... &
\end{eqnarray}
By inspection of the diagrams in Fig.~\ref{fig1}, each term has the following
interpretation. The first term (zeroth power in $V$), is the first diagram in the
left hand side in this figure. Using the Goldstone's rules, the
analytical expression for a one particle-one hole bubble
is given by $\Pi^{0}$. When the bubble is put on it mass shell, we take the imaginary part, ${\cal I}$.
The next contribution (first power in $V$), is represented by the second and third diagrams in
the same figure. In the second (third) diagram the lower (upper) bubble is on the mass
shell. Analytically, the bubble on
it mass shell is given by ${\cal I}$, while the other bubble (in the same diagram) is
off the mass shell, ${\cal R}$. As both contributions
(second and third diagrams) are identical, one has
a factor two (i.e., $2  {\cal R} \; V  \; {\cal I}$).
This association between diagrams and physical states fails for the next order contribution.
There are three contributions, where the first one is shown in Fig.~\ref{fig1},
while the two remainders ones are the same draw, but with the dashed-line (which represents the
configuration on it mass shell), in the middle and upper bubble. The first term
(i.e., $3  {\cal R}^{2} {\cal I}$) is easily interpreted as the sum of these three contributions,
where only one bubble at a time is on the mass shell and a factor three results from the equality of
the three contributions. However, the ${\cal I}^{3}$-term can not be interpreted: in terms of
Eq.~(\ref{stfun2}), all diagrams represent the square of a transition amplitude.
To put it in other words, in one diagram only one configuration can be on the mass shell. The
${\cal I}^{3}$-term would imply a diagram with three bubbles
simultaneously on the mass shell.

We go back to Eq.~(\ref{dyson2}) where  we keep only the
terms compatible with Eq.~(\ref{stfun2}),
\begin{eqnarray}
\label{ring7}
Im(\Pi^{0}) & = & {\cal I} \nonumber \\
Im(\Pi^{0} V  \Pi^{0}) & = & 2  {\cal R} \; V \; {\cal I}\nonumber \\
Im(\Pi^{0}(V \, \Pi^{0})^{2}) & = & 3  {\cal R}^{2}\; V^{2}  \; {\cal I}\nonumber \\
& ... & \nonumber \\
Im(\Pi^{0}(V \, \Pi^{0})^{N}) & = & (N+1) \, {\cal R}^{N} \; V^{N}  \; {\cal I}\nonumber \\
& ... &
\end{eqnarray}
Each term in this series has a straightforward physical interpretation in terms
of the so-called ring diagrams. For this reason, we call the sum as $\Pi^{ring}$.
The summation can be easily performed once we notice that,
\begin{equation}
\label{ring8}
\frac{d}{d ({\cal R} \, V)} {\left ( \frac{1}{1-{\cal R} \, V}
\right )} =1+2 {\cal R} \, V +
3 ({\cal R} \, V)^{2}+ 4 ({\cal R} \, V)^{3}+ \ldots
\end{equation}
The final result for the sum is,
\begin{equation}
\label{stfuncring}
S^{ring}_{C \, (L,T)}(q_0, \v{q}) = \frac{\cal I}
{(1-{{\cal V}_{C \, (L,T)} \, \cal R})^{2}}.
\end{equation}
It should be noted that the expression in the left hand side in Eq.~(\ref{dyson3})
is the sum of the series in the right hand side, as long as $|V \, \Pi^{0}| < 1$,
therefore we have,
\begin{widetext}
\begin{equation}
\label{conver}
|V \, \Pi^{0}| = |V| \, \sqrt{{\cal R}^{2} + {\cal I}^{2}} < 1, \;
as \; {\cal R}, {\cal I} \in Re \Rightarrow |V \, {\cal R}| < 1 \therefore
1-{{\cal V}_{C \, (L,T)} \, \cal R} \neq 0.
\end{equation}
\end{widetext}
Therefore, if the sum in Eq.~(\ref{dyson3}) exists, so does the one in
Eq.~(\ref{stfuncring}). In order to obtained $S^{Dys}$ and $S^{ring}$, only
the imaginary part in Eq.~(\ref{dyson2}) has been evaluated. For completeness,
in Appendix~\ref{append} the real parts are also calculated.

\begin{figure}[h]
\centerline{\includegraphics[scale=0.57]{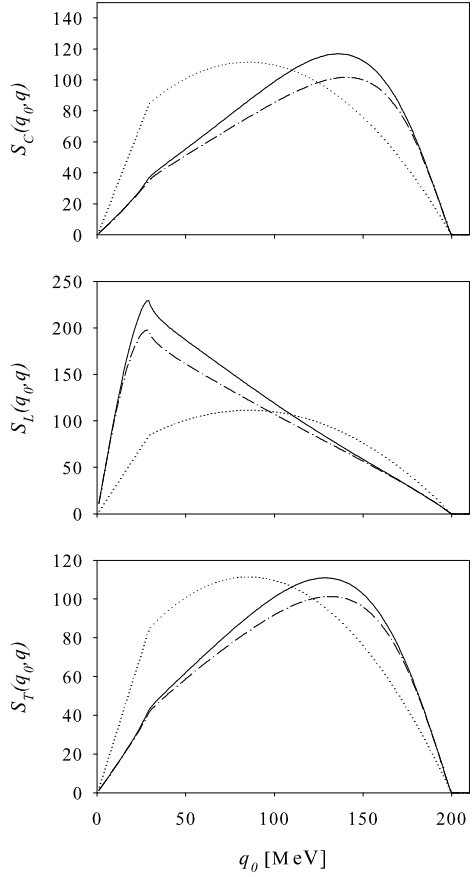}}
\caption{Response function per nuclear volume.
In each graph, the dot line represents $S^{0}$, the
continuous one $S^{ring}$ and the dash-dot $S^{Dys}$.
The momentum transfer by the external operator is chosen
as $q=400$ MeV/c, while the parameters entering in $V$
are $f=0.3$, $g'=0.5$, $\Lambda_{\pi}=1300$~MeV/c and
$\Lambda_{\rho}=1700$~MeV/c. For the Fermi momentum it has been
used, $k_{F}=1.36$ 1/fm.
The functions $S_{C, \, L \, T}$ are given
in units of MeV$^{-1}$ fm$^{-3}$ $\times 10^{-5}$.}
\label{fig2}
\end{figure}

As mentioned, we have two different response functions, $S^{Dys}$ and $S^{ring}$. In
Fig.~\ref{fig2} we have plotted the numerical result for these two functions, just to
show that the difference between them is relevant. It should be emphasized that
both response functions are valid solutions for two different ways of dealing
with the residual interaction $V$: the $S^{Dys}$-response is the solution of the
Dyson's equation and the $S^{ring}$-response is the sum of the ring diagrams.
In the present contribution, it is claimed that the interpretation of the
solution of the Dyson's equation in terms of ring diagrams is wrong, as long as
the polarization insertion $\Pi^{0}$ has a not-null imaginary part. In some physical
problems, such as the study of zero sound~\cite{je80,st83} or core polarization~\cite{os93},
only the real part in Eq.~(\ref{dyson3}) is needed. By inspection of Eqs.~(\ref{dyson4}) and
(\ref{ring10}), it is easy to check that $\Pi^{Dys}=\Pi^{ring}$ when ${\cal I}=0$.
In this case, the solution of the Dyson's equation is also the sum of the ring series.
This element could have been misleading in the former interpretation of the solution
of the Dyson's equation. Before we end this paragraphs, another point should be
addressed: the interpretation of the solution of the Dyson's equation
in term of Eq.~(\ref{stfun2}). This can be done as follows. We
work with the residual interaction,
\begin{equation}
\label{dysonv1}
V^{Dys}(q_0, \v{q}) = V(q) + V(q) \, \Pi^{0}(q_0, \v{q}) \,
V^{Dys}(q_0, \v{q}).
\end{equation}
which has the simple solution,
\begin{equation}
\label{dysonv2}
V^{Dys} = \frac{V}{1 - V \, \Pi^{0}}.
\end{equation}
When this complex interaction is used in replacement of $V$ in the second and
third diagrams in Fig.~\ref{fig1}, a solution of the Dyson's equation
compatible with Eq.~(\ref{stfun2}) is obtained. Analytically,
using Eqs.(\ref{dyson2}) and (\ref{dysonv2}),
it is straightforward to check that,
\begin{equation}
\label{dysonv3}
\Pi^{Dys} = \Pi^{0} \, + \,  \Pi^{0} V^{Dys} \, \Pi^{0}.
\end{equation}
In this case, only three
Goldstone's diagrams comes into play (the first, second and third in
Fig.~\ref{fig1}, with the physical states as marked in this figure).
An expansion in term of ring diagrams of Eq.~(\ref{dysonv2}) is possible, but making
no connection with physical states.

As a further quotation,
the Dyson's equation can be split into it the real and the imaginary part.
In any case, the solution is the one given in Eq.~(\ref{dyson4}).
If $\Pi^{ring}$ (see Eq.~(\ref{ring10})), is replaced in the Dyson's equation, the
imaginary part of this equation is satisfied, but not the real part. This observation
is given only as a warning: in the cases in which the Dyson's equation is solved
numerically, no matters if it is needed only the imaginary part of the solution. Both
real and imaginary parts should be found.

As a concluding remark for this contribution, we have discussed
the response function employing two different ways of dealing with
the residual interaction. The first one is by
using the solution of the Dyson's equation and in the second,
we have analyzed the ring diagrams.
For the ring diagrams, we have taken special care of the configuration
which is on the mass shell and the interpretation of these diagrams in
terms of the Eq.~(\ref{stfun2}). A similar analysis for the solution of the
Dyson's equation has been proposed for completeness. Both analytically and
numerically, these solutions are different. They represent different
approximations and they are both correct. A step forward in this
kind of analysis would be the discussion of the Continued Fraction
Approximation~\cite{sc89}, a subject which has been paid some
attention recently~\cite{ma08}.

\newpage

\begin{acknowledgments}
I would like to thank A. Polls, for fruitful
discussions and for the critical reading of the
manuscript.
This work has been partially supported by the CONICET,
under contract PIP 6159.
\end{acknowledgments}

\appendix
\section{}
\label{append}

The lowest-order polarization insertion is,
\begin{widetext}
\begin{equation}
\label{polpro}
\Pi^{0}(q_0, \v{q}) = 4 \int \frac{d^{3} k}{(2 \pi)^{3}} \,
\theta(|\v{q}+\v{k}|-k_{F}) \theta(k_{F}-k) \,
\left( \frac{1}{q_0- t_{q+k}+t_{k} + \imath \eta} -
\frac{1}{q_0+ t_{q+k}-t_{k} - \imath \eta}
\right)
\end{equation}
\end{widetext}
where $t_p=p^{2}/(2 m)$, with $m$ being the nucleon mass. In this
equation $k_F$ is the Fermi momentum.

We present now our model for the residual interaction $V$,
\begin{widetext}
\begin{equation}
\label{inter}
V(q) = \frac{f_{\pi}^2} {\mu_{\pi}^2} \Gamma_{\pi}^2 (q)
(\,f \, + \, g' \,
\v{\sigma} \cdot \v{\sigma'} \,
\v{\tau} \cdot \v{\tau'} \; + \;
V_{\pi}(q) \v{\sigma} \cdot \v{\widehat{q}} \,
\v{\sigma'} \cdot \v{\widehat{q}} \,
\v{\tau} \cdot \v{\tau'}
\; + \; V_{\rho}(q)
(\v{\sigma} \times \v{\widehat{q}}) \cdot
(\v{\sigma'} \times \v{\widehat{q}}) \,
\v{\tau} \cdot \v{\tau'}),
\end{equation}
\end{widetext}
where it has been taken the static limit. Therefore,
$V_{\pi}(q) =  -q^2/(q^2 + \mu_{\pi}^2)$ and
$V_{\rho}(q) = -(\Gamma_{\rho}/
\Gamma_{\pi})^{2} \, C_{\rho} \,
q^2/(q^2 + \mu_{\rho}^2)$,
where $\mu_{\pi}$ ($\mu_{\rho}$ )
is the pion (rho) rest mass, $f_{\pi}^2/4 \pi=0.081$
and $C_{\rho} = 2.18$.
The form factor of the $\pi NN$ ($\rho NN$) vertex
is $\Gamma_{\pi}$ ($\Gamma_{\rho}$),
where $\Gamma_{j}=
((\Lambda^{2}_{j}-m^{2}_{j})/(\Lambda^{2}_{j}+q^{2}))^{2}$.
Using the property,
\begin{equation}
\label{sigma}
\v{\sigma} \cdot \v{\sigma'} =
\v{\sigma} \cdot \v{\widehat{q}} \,
\v{\sigma'} \cdot \v{\widehat{q}} \, +
(\v{\sigma} \times \v{\widehat{q}}) \cdot
(\v{\sigma'} \times \v{\widehat{q}}),
\end{equation}
the Eq.~(\ref{inter}) can be rewritten as,
\begin{widetext}
\begin{equation}
\label{inter2}
V(q) = \frac{f_{\pi}^2} {\mu_{\pi}^2} \Gamma_{\pi}^2 (q)
({\cal V}_C \; + \; {\cal V}_L \v{\sigma} \cdot \v{\widehat{q}} \,
\v{\sigma'} \cdot \v{\widehat{q}} \,
\v{\tau} \cdot \v{\tau'}
\; + \; {\cal V}_T
(\v{\sigma} \times \v{\widehat{q}}) \cdot
(\v{\sigma'} \times \v{\widehat{q}}) \,
\v{\tau} \cdot \v{\tau'}),
\end{equation}
\end{widetext}
with obvious definitions for ${\cal V}_{C, \, L, \, T}$.

As a final point for this Appendix, we split the solution of the
Dyson's equation (Eq.~(\ref{dyson})), into it real and imaginary parts,
\begin{equation}
\label{dyson4}
\Pi^{Dys} = \frac{{\cal R}(1-V \, {\cal R})- V \, {\cal I}^{2}}
{(1-V \, {\cal R})^{2}+ (V \, {\cal I})^{2}} \; + \;
\frac{{\cal I}}
{(1-V \, {\cal R})^{2}+ (V \, {\cal I})^{2}} \, \imath
\end{equation}
We now perform the same procedure as in Eq.~(\ref{ring7}), but for the
real part of the ring series,
\begin{eqnarray}
\label{ring7r}
Re(\Pi^{0}) & = & {\cal R} \nonumber \\
Re(\Pi^{0} V  \Pi^{0}) & = & {\cal R}^{2} \; V \nonumber \\
Re(\Pi^{0}(V \, \Pi^{0})^{2}) & = & {\cal R}^{3} \; V^{2} \nonumber \\
& ... & \nonumber \\
Re(\Pi^{0}(V \, \Pi^{0})^{N}) & = & {\cal R}^{N+1} \; V^{N} \nonumber \\
& ... &
\end{eqnarray}
where the sum is,
\begin{equation}
\label{ring9}
Re (\Pi^{ring}) = \frac{{\cal R}}{1-V \, {\cal R}}.
\end{equation}
Finally, we can write,
\begin{equation}
\label{ring10}
\Pi^{ring} = \frac{{\cal R}(1-V \, {\cal R})}
{(1-V \, {\cal R})^{2}}  \; + \;
\frac{{\cal I}}
{(1-V \, {\cal R})^{2}} \, \imath
\end{equation}

\end{document}